\documentclass[12pt]{article}

\usepackage{amsmath,amssymb}
\usepackage[pdftex]{graphicx}
\usepackage{url}
\usepackage{epsfig}
\usepackage{color}
\usepackage{verbatim}
\usepackage{enumitem}

\newcommand{\Vb}{V_b}
\newcommand{\Vr}{V_r}
\newcommand{\RS}{\mathit{RS}}
\newcommand{\AS}{\mathit{AS}}

\newcommand{\Aggr}{{\mathsf{Aggr}}}
\newcommand{\Detect}{\mathsf{Detect}}
\newcommand{\TimeDiscount}{{\mathsf{TimeDiscount}}}
\newcommand{\MEA}{\mathsf{MEA}}
\newcommand{\TOA}{\mathsf{TOA}}

\newcommand{\GS}{{\mathsf{GS}}}
\newcommand{\GKeyGen}{{\mathsf{GKeyGen}}}
\newcommand{\GSign}{{\mathsf{GSign}}}
\newcommand{\GVerify}{{\mathsf{GVerify}}}
\newcommand{\GJoin}{\mathsf{GJoin}}
\newcommand{\Open}{{\mathsf{Open}}}

\newcommand{\T}{\mathbb{T}}

\newcommand{\BBS}{{\mathsf{BBS}}}
\newcommand{\BKeyGen}{{\mathsf{BKeyGen}}}
\newcommand{\BSign}{{\mathsf{BSign}}}
\newcommand{\BVerify}{{\mathsf{BVerify}}}
\newcommand{\BJoin}{\mathsf{BJoin}}
\newcommand{\BOpen}{{\mathsf{BOpen}}}
\newcommand{\BUpdate}{{\mathsf{BUpdate}}}
\newcommand{\asn}{{\leftarrow}}

\newcommand{\gpk}{\mathsf{gpk}}
\newcommand{\gsk}{\mathsf{gsk}}
\newcommand{\gmsk}{\mathsf{gmsk}}

\newcommand{\msg}{\mathsf{msg}}
\newcommand{\fr}{\mathsf{fr}}

\newcommand{\G}{{\mathbb G}}

\newcommand{\rcert}{{\mathsf {rcert}}}
\newcommand{\calR}{{\mathcal{R}}}
\newcommand{\Z}{{\mathbb Z}}

\begin{document}

\title{Private reputation retrieval in public - a privacy-aware
  announcement scheme for VANETs
  \footnote{The initial idea of this material was published in the 
WiVeC proceedings \cite{wivec} and a comprehensive solution was presented 
in CTTD 2013 (no proceedings) \cite{CTTD}. This paper is a full version 
of the whole work.}}

\author{
Liqun Chen, Hewlett Packard Labs, Bristol, BS34 8QZ, UK.\\
Email: liqun.chen@hpe.com. 
\and
Qin Li, Email: Qin.Li.2008@live.rhul.ac.uk.\\
\and
Keith~M.~Martin, Siaw-Lynn Ng, Information Security Group, \\ Royal Holloway, 
University of London, Egham, TW20 0EX, UK. \\
Email: \{keith.martin, s.ng\}@rhul.ac.uk
}
\date{\today}
\maketitle

\begin{abstract}
An announcement scheme is a system that facilitates vehicles to
broadcast road-related information in vehicular \emph{ad hoc} networks
(VANETs) in order to improve road safety and efficiency.  Here we
propose a new cryptographic primitive for public updating of
reputation score based on the Boneh-Boyen-Shacham short group
signature scheme.  This allows private reputation score retrieval
without a secure channel. Using this we devise a privacy-aware
announcement scheme using reputation systems which is reliable,
auditable and robust.  
\end{abstract}

\section{Introduction}
\label{sec:intro}

Vehicular \emph{ad hoc} networks (VANETs) allow vehicles to exchange
information about vehicle, road, and traffic conditions. We call a
system that facilitates vehicles to exchange road-related information
an \textit{announcement scheme}. If the road-related information
exchanged in an announcement scheme is reliable then this would enable
a safer and more efficient travelling environment. We say that a
message is \textit{reliable} if it reflects reality. Unreliable
messages may result in various consequences, for example journey
delays or accidents. Unreliable messages may be a result of vehicle
hardware malfunction. For example, if a sensor in a vehicle is faulty
then messages generated from the faulty sensor may be
false. Unreliable messages can also be generated intentionally. For
example, some vehicles may generate and broadcast false road
congestion messages with the intention to deceive other vehicles into
avoiding certain routes. In extreme cases, unreliable message may lead
to injuries and even deaths. Hence, an announcement should have the
following functionalities:

\begin{itemize}
\item \textit{Message reliability evaluation}. Vehicles should be able to
  evaluate the reliability of received messages.
\item \textit{Auditability}. Vehicles that broadcast unreliable
  messages should be identified and revoked.
\end{itemize}

In addition, the announcement scheme should satisfy the following
security requirements:

\begin{itemize}
\item \textit{Robustness}. The accuracy of message reliability
  evaluation and auditability should not be affected by attacks, from
  both internal and external adversaries.
\item \textit{Privacy awareness}. The privacy of vehicles should be
  protected, since the information about vehicle position is often
  sensitive to vehicle users. The vehicle privacy has two facets as
  follows:
        \begin{itemize}
        \item \textit{Anonymity}. The identity of a vehicle should not
          be revealed from data broadcast by the vehicle.
        \item \textit{Unlinkability}. Multiple pieces of data
          broadcast by the same vehicle should not be linked to each
          other.
        \end{itemize}
\end{itemize}

In \cite{wivec} a privacy-aware reputation-based announcement scheme
for VANETs was proposed.  This scheme relies on a centralised
reputation system with an off-line trusted authority, and uses group
signatures to allow vehicles to make authenticated announcements
anonymously.  An announcement will be accepted as reliable if the
announcing vehicle has a sufficiently high reputation.  The reputation
reflects the extent to which the vehicle has announced reliable
messages in the past. It is computed and updated based on
\textit{feedback} reported by other vehicles. The reputation scores of
all vehicles are managed by a central \textit{reputation server}.
This scheme has two fundamental weaknesses: firstly, the decision as
to whether an announcement is trustworthy or not is made by the
reputation server rather than the receiving vehicle, since only
vehicles deemed reputable by the reputation server are given signing
keys, and the signatures do not reveal what the reputation scores are.
Secondly, a secure channel is required for the retrieval of new
signing keys (and hence new reputation status).  In \cite{wivec} a
brief sketch was provided to indicate how these weaknesses may be
overcome.  Here we describe in full a new cryptographic primitive
which enables the design of a scheme to address these two weaknesses:

\begin{enumerate}
\item We propose a new tool for public updating of reputation score
  based on the Boneh-Boyen-Shacham (BBS) short group signature scheme
  \cite{boneh2004short}.  When the reputation score of a group member
  $\Vb$ changes, $\Vb$ is able to update its signing key using a
  public value in such a way that its signature is bound to the new
  reputation score.  This signature can be verified by other group
  members, again using a public value.  This overcomes the significant
  problem of having to establish a secure channel for reputation score
  retrieval.

\item Using this new cryptographic primitive we improve the scheme of
  \cite{wivec} to support flexible decision-making on the part of the
  receiving ​vehicle.  If a reputation score is visible in a group
  signature then a receiving vehicle may decide whether to trust the
  announcement depending on the type of announcement and the
  announcing vehicle's reputation score.  Our scheme here supports
  this.
\end{enumerate}

\section{Related Work}
\label{sec:related}

There have been a number of announcement schemes proposed to evaluate
the reliability of messages in VANETs. These can be categorised into
two main groups: \textit{threshold method} and
\textit{reputation-based method}.

A majority of announcement schemes,
e.g.~\cite{CNW09,DDSV09,KWL09,RAH06, wu2010balanced,CPAV,NoPairings}, use the
threshold method: a message is believed reliable if it has been
announced by multiple distinct vehicles whose number exceeds a
threshold within a time interval. This method gives rise to the
problem of \emph{distinguishability of message
  origin}~\cite{DBLP:conf/iptps/Douceur02} - how to tell if two
messages are made by two distinct vehicles if vehicles are anonymous
and their activities are unlinkable. Solutions to this problem include
using message linked group signatures~\cite{wu2010balanced} and a
combination of Direct Anonymous Attestation \cite{ChenDAA} and 
1-time anonymous authentication
\cite{teranishi2004k}.   
In addition, this method is only suitable for
event-driven messages, where multiple vehicles may
broadcast the same message. It is not suitable for beacon messages,
where a beacon is only broadcast by one vehicle.

There have been several reputation-based methods, such
as~\cite{DFM05,MJT10,schmidt2008vehicle,Li2012reputation,wivec,central}. The
schemes in~\cite{DFM05,MJT10,schmidt2008vehicle} adopt a decentralised 
infrastructure while those in~\cite{Li2012reputation,wivec,central} use 
a centralised system. In \cite{Li2012reputation}
Li et al. proposed a reputation-based announcement scheme that aims to
provide message reliability evaluation, auditability, and
robustness. A vehicle periodically retrieves its \textit{reputation
  certificate}, which contains its reputation score, from the central
authority. When a vehicle broadcasts a message, it attaches its
reputation certificate to the message. A receiving vehicle extracts
the reputation score and then infers the reliability of the message. A
vehicle whose reputation score decreases beyond a threshold is revoked
by the central authority. This is achieved by no longer providing the
vehicle its reputation certificate in the future.  However, this
scheme~\cite{Li2012reputation} lacks the provision of privacy
protection to vehicles: messages and feedback are linkable and not
anonymous. An adversary is able to conduct a \textit{profiling attack}
to learn the moving trace of a target vehicle. This drawback may
affect the willingness of vehicles to participate in the announcement
scheme.  The scheme in \cite{central} suffers from the same drawback.
This drawback is rectified in the scheme of \cite{wivec}, which we
will describe in detail in Section~\ref{sec:wivec}.  On the other
hand, \cite{multihop} considers how a reputation-based scheme may be
extended to allow multihop communications.

In~\cite{DFM05}, upon receiving a
message, a vehicle can append its own opinion about its reliability to
the message. This message is then forwarded, along with the appended
opinion. In this scheme, a vehicle verifies the reliability of a
message by aggregating all the opinions appended to the
message. However, its robustness against possible collusion of
adversaries is not addressed. Vehicle privacy is also not provided by
this scheme. Besides, receiving vehicles have to bear a heavy
computational burden in order to verify the digital signature signed
on each opinion -  every vehicle has to verify many signatures
before appending its own. In addition, implementation details, such as
initialisation and malicious vehicle revocation, are not discussed.

In \cite{MJT10}, the reliability of a message is evaluated according
to three different types of trust value regarding the message
generating vehicle: \textit{role-based trust},
\textit{experience-based trust} and \textit{majority-based
  trust}. Role-based trust assumes that a vehicle with a certain
predefined role, such as traffic patrol or law enforcing authorities,
has a high trust value. Majority-based trust is similar to the
threshold method that we discussed earlier. Experience-based trust is
established based on direct interactions: a vehicle trusts another
vehicle if it has received many reliable messages from the other
vehicle in the past. A similar approach to experienced-based trust was
also proposed in~\cite{PJF06}. A drawback of this approach is that it
requires vehicles to establish a long-term relationship with each
other, which may not be practical in a large VANET
environment. Furthermore, it also requires vehicles to store
information regarding vehicles that they have encountered in the
past. This may lead to not just a demand for storage but also a demand
for rapid searching through the information to make a decision which
may result in a lag in responding to potentially critical
events.. Lastly, robustness and vehicle privacy are not provided.

In \cite{schmidt2008vehicle}, a vehicle conducts behaviour analysis
about another vehicle based on some observable information about the
target vehicle, such as its positions, movements and messages
broadcast in the past. The result of this analysis is used to
determine the trustworthiness of the target vehicle and the
reliability of messages broadcast by it. However, in this scheme,
vehicles have to make observations before making a decision, which may
not be feasible in VANETs. In addition, robustness and vehicle privacy
are not provided by this scheme.  

Compared with existing threshold and reputation-based schemes, the
schemes~\cite{Li2012reputation,wivec} feature the following:
\begin{itemize}
\item They enable immediate evaluation: a receiving vehicle does not
  require multiple messages in order to verify the reliability of a
  message.
\item They support reliability evaluation of both beacon and
  event-driven messages.
\item They support revocation of maliciously-behaving vehicles.
\item They provide strong robustness against external adversaries, and
  robustness against internal adversaries to a reasonably good level.
\item They achieve a good level of efficiency.
\end{itemize}

In addition to the features above, the
scheme~\cite{wivec} also provides a good level of vehicle privacy.

\section{Privacy-aware reputation-based announcement
scheme}
\label{sec:wivec}

For completeness, we include a brief description of the privacy-aware
reputation-based announcement scheme \cite{wivec}. We describe first
the algorithms and protocols that are required:
\begin{itemize}
\item A secure and privacy-aware mutual entity authentication protocol
  $\MEA^+$. We use $\MEA^+\{A \rightarrow B : m\}$ to denote the
  situation where the message $m$ is sent from $A$ to $B$ where both
  communicating parties $A$ and $B$ are assured of: 1) the identity of
  each other, 2) the freshness of the communication, and 3) the
  protection of the communication against all entites (apart from $A$
  and $B$) with respect to anonymity and unlinkability.  This protocol
  will be used by vehicles to retrieve their reputation and report
  feedback.  It can be instantiated by using a secure probabilistic
  encryption scheme to establish an encrypted channel, and then
  executing a suitable authentication protocol in the encrypted
  channel.

\item A secure and privacy-aware \textit{two-origin authentication}
  protocol $\TOA^+$. We use $\TOA^+\{A: m_1, m_2: C \}$ to denote the
  situation where the message $(m_1, m_2)$ is broadcast by $A$, and 
  a recipient is given the assurance
  that: 1) $m_1$ originates from a legitimate (but unidentified) entity,
  2) $m_2$ originates from a third
  party C, and 3) $m_2$ is bound to messages originating from $A$.
  This protocol will be used by
  vehicles to broadcast messages.  It can be implemented
  using, for example, a group signature scheme.

\item An aggregation algorithm $\Aggr$, which will be used to
  aggregate feedback and produce reputation scores for vehicles.

\item A data analysis algorithm $\Detect$, which will be used to
  detect malicious vehicles based on feedback.

\item A time discount function $\mathsf{TimeDiscount}$. This is a
  non-increasing function whose range is $[0, 1]$. It takes as input a
  non-negative value representing a time difference, and outputs a
  number between 0 and 1.
     One simple example is:
\[
\mathsf{TimeDiscount}(t) = \left\{
               \begin{array}{l l}
                       1 - t/ \Psi_{td}    &  \text{if } t < \Psi_{td}; \\
                       0                 &  \text{if } t \ge \Psi_{td}, \\
               \end{array} \right. \]

The TimeDiscount function is used to determine the
freshness of a vehicle's reputation score in order to
prevent abuse of the system. For instance, a vehicle may
continue to announce messages using its old reputation credential with
higher reputation score in order to avoid retrieving its latest
reputation credentials that may have lower reputation score after
misbehavior. The TimeDiscount function makes sure that the reputation score is
``discounted'' with time.

In this case we take the absolute value of the difference between the
current time when a message is received and the time the reputation
certificate was retrieved.  An older reputation certificate gives a
larger difference in value which results in a lower value of
discounted reputation score.  This is by no means the only possibility
for time discount functions but we have chosen this as the most
straightforward option.

\item A threshold $\Psi$ between 0 and 1, which will be used to
  determine whether a reputation score is sufficiently high.
\end{itemize}

For completion we will introduce notation for a group signature
scheme that will be used to implement $\TOA^+$ in
\cite{wivec}:

A secure \textit{group signature scheme}~\cite{chaum1991group,
  ateniese2000practical, boneh2004short}, denoted by $\GS =
(\GKeyGen$, $\GJoin$, $\GSign$, $\GVerify$, $\Open)$ where $\GKeyGen$,
$\GJoin$, $\GSign$, $\GVerify$ and $\Open$ denote group public key
generation, group member secret key generation, group member signing,
group verification, and signer revealing algorithms, respectively. 
All members of the group has access to the group public key while each
individual member is given its own group member secret key.  A group
signature scheme is a digital signature scheme that has the following
properties:
\begin{itemize}
\item Each group member can sign messages (using its group member secret key).
\item A receiver can verify whether the signature was signed by a
  group member (using the group public key with the group verification
  algorithm), but cannot discover which group member signed it.
\item Any two messages signed by a group member cannot be linked.
\item A signature can be ``opened'' by a group manager (using the
  signer revealing algorithm), if necessary, so that the group member
  who signed the message is revealed.
\end{itemize}

(Note that we treat the entire system as one group.  Members join when
they register and leave when they are revoked and these are all
controlled centrally.  Keys are only updated by time.  There is no
``group'' in the sense of dynamic networks where members may join and
leave different groups at will.  There are indeed some work (for
example, \cite{groupformation,groupleader}) where vehicles travelling
in a certain direction and locality form groups and communicate with
each other within the group.  That would happen \emph{within} our
framework.)

\subsection{Description of the scheme}
\label{sub:wivec_description}

This scheme has a centralised architecture with off-line central
entities - we have taken the centralised approach since there is generally a
centrally authority governing the registration and administration of
vehicles. \textit{Vehicles} ($V$s) are the end users. We assume that $V$s are
mobile entities that have computational and short range wireless
communication devices. The functionalities of vehicles include:

\begin{enumerate}
\item generating and broadcasting messages to neighbouring vehicles,
\item receiving messages from neighbouring vehicles and evaluating their
reliability, and
\item reporting feedback.
\end{enumerate}

There are two logical off-line central entities: a \textit{reputation
  server} ($\RS$), and an \textit{administrative server} ($\AS$). The
$\RS$ computes \textit{reputation scores} for vehicles based on
\textit{feedback} reported by vehicles. The functionality of the $\AS$
includes:

\begin{enumerate}
\item admitting new vehicles into the system and revoking malicious
  vehicles from the system,
\item providing reputation endorsement for vehicles, and
\item collecting feedback reported by
vehicles.
\end{enumerate}

The $\AS$ has multiple remote wireless communication interfaces so
that vehicles can intermittently communicate with the $\AS$ in a
convenient and frequent manner (for example once a day). Note that we
do not require a vehicle to be able to constantly communicate with the
$\AS$, meaning that the $\RS$ and $\AS$ are off-line entities. We
assume that the $\RS$ and $\AS$ are trusted and interact honestly with
each other, and the communication channel between the $\RS$ and $\AS$
is secure (authenticated, confidential, and integrity protected). We
assume that the $\AS$ has a clock. We also assume that a vehicle has a
clock that is loosely synchronised with the clock of the
$\AS$. Although the $\RS$ and $\AS$ can be separately distributed, one
convenient setting during an implementation is to make them form a
single trusted entity, a \textit{central authority}. We also assume
that the communication channels between the $\AS$ and vehicles, and
those between vehicles are publicly open, and thus subject to attacks.

\begin{enumerate}[label=(\Roman*)]
\item \textit{Scheme Initialisation}.
\begin{enumerate}
\item The $\AS$ regulates its clock, and deploys its remote wireless
  communication interfaces.

\item The $\RS$ creates a database, and installs $\Aggr$ and
  $\Detect$.

\item The $\AS$ installs $\GS$, $\MEA^+$,
  $\TimeDiscount$, and $\Psi$, and initialises the cryptographic keys
  to be used by $\AS$ during future execution of $\MEA^+$.
\item The $\AS$ divides the time into time intervals $(\T_{0}, \T_{1},
  \T_{2}, \cdots)$. The length of a time interval is configurable. For
  example, each time interval can be one day. For each time interval
  $\T_{i}$, $\AS$ uses $\GKeyGen$ to generate a group public
  key $pk_{i}$ and uses $\GJoin$ to generate a set of
  corresponding group member secret keys $(sk^{1}_{i}, sk^{2}_{i},
  \cdots, sk^{n}_{i})$ where $n$ is the number of vehicles in the
  system. A secret key $sk^{j}_{i}$ is to be used by vehicle $V_{j}$
  during the time interval $\T_{i}$. Group member secret keys
  $(sk^{j}_{0}, sk^{j}_{1}, sk^{j}_{2}, \cdots)$ are to be used by
  $V_{j}$ during the corresponding time intervals $(\T_{0}, \T_{1},
  \T_{2}, \cdots)$. The keys $sk^{j}_{i}$ for all $i$ and $j$ are kept
  confidential for future use.
\end{enumerate}

\item  \textit{Vehicle Registration}.
\begin{enumerate}
\item The $\AS$ initialises the cryptographic keys to be used by $V$
  during future execution of $\MEA^+$.
\item The $\AS$ provides $V$ with $\MEA^+$, $\mathsf{GSign}$,
  $\mathsf{GVerify}$, the keys generated from the previous step, and
  $(pk_{0}, pk_{1}, pk_{2}, \cdots)$. We assume that this is conducted
  over a secure channel.
\item The $\AS$ requests the $\RS$ to create a record in its database for vehicle $V$.
\end{enumerate}

\item \textit{Reputation Retrieval}.  When a vehicle $V_{b}$
drives into the proximity of a wireless communication interface during
a time interval $\T_{i}$, whose beginning time is denoted by $t_{i}$,
it retrieves its reputation information as follows:
\begin{enumerate}
\item $\Vb$ and the $\AS$ execute $\MEA^+$ to
  establish an encrypted and mutually authenticated channel.
\item Upon retrieving $(r,\Vb, t_i)$, the reputation score $r$ of $\Vb$ at
  the current time $t_i$, from the $\RS$, the $\AS$ computes $\Vb$'s
  time discounted reputation scores $(r'_{i}, r'_{i+1}, \cdots,
  r'_{i+m})$ until $r'_{i+m+1} < \Psi_{r}$. A time discounted
  reputation score $r'_{i+k} = r \cdot \mathsf{TimeDiscount}(t_{i+k} -
  t_{i})$, where $t_{i}$ and $t_{i+k}$ denote the beginning times of
  $\T_{i}$ and $\T_{i+k}$, respectively. These scores correspond to
  the time intervals $(\T_{i}$, $\T_{i+1}$, $\cdots$, $\T_{i+m})$,
  respectively. Note that $r'_{i+k} \ge \Psi_{r}$ for $0 \le k \le m$
  and $r'_{i+k} < \Psi_{r}$ for $k > m$. In other words, $V_{b}$ is
  considered as \textit{reputable} for the time intervals $\T_{i},
  \cdots, \T_{i+m}$.
\item The $\AS$ sends $\Vb$ in the encrypted and mutually
  authenticated channel the group member secret keys
  $(sk^{b}_{i},\cdots, sk^{b}_{i+m})$, which correspond to $\T_{i},
  \cdots, \T_{i+m}$.
\end{enumerate}

\item \textit{Message Broadcast}.
A message $m$ is broadcast by $\Vb$ as follows:
\begin{enumerate}
\item $\Vb$ retrieves the current time from its clock and identifies
  its corresponding time interval, say $\T_{i}$.
\item $V_{b}$ uses $\GSign$ and $sk^{b}_{i}$ that corresponds to the
  time interval $\T_{i}$, to generate a signature $\theta$ on $(m,i)$, and
  forms a \textit{message tuple} $M = (m, i, \theta)$. $\Vb$ then
  broadcasts $M$ to its neighbouring vehicles.
\item Upon receiving $M$, a receiving vehicle $\Vr$ immediately
  identifies the current time interval $\T_{j}$ from its clock.  $\Vr$
  checks if $j = i$. If so then $\Vr$ uses $\GVerify$ and $pk_{i}$,
  which corresponds to $\T_{i}$, to verify $\theta$. Upon successful
  verification, $\Vr$ considers $\Vb$ to be reputable, and the message
  $m$ to be reliable. The message tuple $M$ is stored for future
  possible feedback reporting. If $j \not= i$ or the verification
  fails then $\Vr$ does not consider $V_{b}$ to be reputable, and
  discards $M$.
\end{enumerate}

\item \textit{Feedback reporting}. When $V_{r}$ has experience
about the event described by message $m$, it is able to judge the
reliability of $m$. Then $V_{r}$ can voluntarily report feedback as
follows:

\begin{enumerate}
\item $V_{r}$ assigns a feedback $f$ based on its experience about the
  reliability of $m$;
\item When $V_{r}$ drives into the proximity of a wireless
  communication interface, $\Vr$ and the $\AS$ execute $\MEA^{+}$ to
  establish an encrypted and mutually authenticated channel, and $\Vr$
  sends $f,M$ to the $\AS$ via the channel.
\item The $\AS$ uses $\Open$ and $pk_{i}$ to open $M$, in order to
  retrieve signer $\Vb$, and sends the $\RS$ the tuple $(f, \Vb,
  \Vr)$.  The $\RS$ stores it in the database.
\item The $\RS$ uses $\Aggr$ and all feedback stored in the database
  to update the reputation of $\Vb$.
\end{enumerate}

\item \textit{Vehicle Revocation}.
The $\AS$ revokes the identified
malicious vehicle by no longer providing them with new group member
secret keys in the future.
\end{enumerate}

In this scheme, a reputation credential of $\Vb$ at time interval
$\T_i$ is represented by a group member secret key $sk^{b}_i$.  Hence
$\TOA^+$ is realised by $\GS$: $\TOA^{+} \{ \Vb : m, (r'_{i}\ge \Psi)
: \AS \} =(m, i, \theta)$, where $\theta =
\GSign_{sk_{i}^{b}}(m,i))$. This gives a recipient assurance that $m$
originated from a reputable (but unidentified) vehicle.

\subsection{Privacy and Robustness}
\label{sub:priv}

This scheme is robust against both external and internal
adversaries with respect to both message fraud (an adversary deceives
a vehicle into believing that a false message is reliable) and
reputation manipulation (an adversary unfairly inflates or deflates
the reputation score of a target vehicle) attacks. It also provides
privacy protection (anonymity and unlinkability) for vehicles against
all adversaries except for the central authority \cite{Li2012reputation,wivec}.

\subsection{Extending to multiple reputation levels}
\label{sub:ext}

As described in Section \ref{sec:intro}, we will extend this scheme to
support multiple reputation levels, thus allowing flexible
decision-making for individual vehicles.  We will also remove the
constraint of having to use a secure channel for credential retrieval.
This extended scheme will be described in Section
\ref{sec:multiple_description}.  Before that we will describe in
Section \ref{sec:modifiedBBS} a novel modification of a group signature
scheme which will underpin our new scheme.

\section{An extension of the BBS scheme}
\label{sec:modifiedBBS}

Here we will describe a modification of the BBS \cite{boneh2004short}
group signature scheme - in essence, both $\MEA^+$ and $\TOA^+$ will
be implemented using this scheme. This will also allow private
reputation score retrieval via a public channel.  While this modified
primitive is designed for application within the scenario of this
paper, it has the potential to be of independent interest.

\subsection{The BBS Scheme}
\label{sub:BBS}

We first briefly describe, informally, the original BBS \cite{boneh2004short}
group signagure scheme.  Formal details and security proofs can be found in
\cite{boneh2004short}.
Let $\G_1$, $\G_2$ and $\G_3$ be three multiplicative cyclic groups of
large prime order $p$.  Let $g_1$ be a generator of $\G_1$ and $g_2$ a
generator of $\G_2$.  Let $\psi$ be a computatble isomorphism from
$\G_2$ to $\G_1$, with $\psi(g_2)=g_1$.  (It is noted in
\cite{boneh2004short} that $\psi$ is needed only for proofs of
security.  We need only to assume that it exists and is efficiently
computable.)

Let $\hat{t}: \G_1 \times \G_2 \rightarrow \G_3$
be a computable bilinear map:
\begin{align*}
& \hat{t}(u^a, v^b) = \hat{t}(u, v)^{ab}
  \;\; \forall u \in \G_1, \; v \in \G_2 \mbox{ and } a, b \in \Z \\
& \hat{t}(g_1,g_2) \neq 1
\end{align*}

We require that the \emph{$q$-Strong Diffie-Hellman} ($q$-SDH) problem
is hard in $(\G_1, \G_2)$ and the \emph{Decision Linear
  Diffie-Hellman} problem is hard in $\G_1$:

The $q$-SDH problem in $(\G_1, \G_2)$ is as follows: given a
$(q+2)$-tuple $(g_1,g_2, g_2^{\gamma},g_2^{\gamma^2 }, \ldots,
g_2^{\gamma^q})$ as input, output a pair
$(g_1^{\frac{1}{\gamma+x}},x)$, where $x \in \Z_p^*$.

The Decision Linear Diffie-Hellman problem is as follows: given $u$,
$v$, $h$, $u^a$, $v^b$, $h^c \in \G_1$ as input, decide whether
$a+b=c$.

The BBS group signature scheme $\BBS = (\BKeyGen$, $\BJoin$, $\BSign$,
$\BVerify$, $\BOpen)$ where $\BKeyGen$, $\BJoin$, $\BSign$, $\BVerify$
and $\BOpen$ denote group public key generation, group member secret
key generation, group member signing, group verification, and signer
revealing algorithms, respectively, is as follows.  (We will write $x
\asn S$ to denote the action of sampling an element from $S$ uniformly
at random and assigning the result to the variable $x$.)

\begin{itemize}
\item $\BKeyGen$:

In key generation $\BKeyGen$ generates $\G_1$, $\G_2$, $\G_3$, $g_1$,
$g_2$, $\psi$ and $\hat{t}$ as described above.  Let $\eta_1,\eta_2
\asn \Z_p^*$, $h \asn \G_1 \setminus \{1_{\G_1}\}$, and set $u, v \in
\G_1$ such that $u^{\eta_1}=v^{\eta_2} = h$.  Let $\gamma \asn
\Z_p^*$, and set $w= g_2^{\gamma} \in \G_2$.

The group public key $\gpk$ will be  $(g_1, g_2, u,v,h, w)$.

The secret key of the group manager is $\gmsk = (\gamma, \eta_1, \eta_2)$.
Note that $(\eta_1, \eta_2)$ is used to open signatures.

Let $H$ be a hash function $H: \{0,1\}^* \to \Z_p$.

\item $\BJoin(b, \gmsk)$:

Each group member $b$ is given a secret key $\gsk_b=(A_b, x_b)$,
where $x_b \asn \Z_p^*$, and $A_b = g_1^{\frac{1}{\gamma+x_b}} \in \G_1$.

\item $\BSign(M, \gsk_b, \gpk)$:

For group member $b$ to sign the message $M$ using $\gpk=(g_1, g_2,
u,v,h, w)$ and $\gsk_b=(A_b, x_b)$, let $\alpha, \beta \asn \Z_p$, and
compute $T_1 = u^{\alpha}$, $T_2=v^{\beta}$, $T_3 = A_b
h^{\alpha+\beta}$.

Now let $r_{\alpha}$, $r_{\beta}$, $r_{x}$, $r_{\delta_1}$, $r_{\delta_2}
\asn \Z_p$, and compute
$R_1 = u^{r_{\alpha}}$,
$R_2 = v^{r_{\beta}}$,
$R_4=T_1^{r_x}u^{-r_{\delta_1}}$,
$R_5 = T_2^{r_x}v^{-r_{\delta_2}}$ and
\[ R_3 = \hat{t}(T_3,g_2)^{r_x}\hat{t}(h, w)^{-r_{\alpha}-r_{\beta}}
\hat{t}(h,g_2)^{-r_{\delta_1}-r_{\delta_2}}.\]

Compute $c=H(M, T_1, T_2, T_3, R_1, R_2, R_3, R_4, R_5)$, and let
$\delta_1 = x_b \alpha$, $\delta_2 = x_b \beta$.  Compute
$s_{\alpha}=r_{\alpha} + c \alpha $,
$s_{\beta}= r_{\beta} + c \beta$,
$s_{x}= r_x + cx_b $,
$s_{\delta_1}= r_{\delta_1} + c \delta_1$ and
$s_{\delta_2}= r_{\delta_2} + c \delta_2$.

The signature on $M$ is $\sigma = (T_1,T_2,T_3, c, s_{\alpha}, s_{\beta}, 
s_{x},$ $s_{\delta_1}, s_{\delta_2})$.

\item $\BVerify(M, \sigma, \gpk)$:

To verify a signature $\sigma = (T_1,T_2,T_3,$ $c, s_{\alpha},
s_{\beta}, s_{x},$ $s_{\delta_1}, s_{\delta_2})$ on the message $M$
using the group public key $\gpk=(g_1, g_2, u,v,h, w)$, compute
$\tilde{R_1}=u^{s_{\alpha}} T_1^{-c}$,
$\tilde{R_2}=v^{s_{\beta}}T_2^{-c}$,
$\tilde{R_4}=T_1^{s_x} u^{-s_{\delta_1}}$,
$\tilde{R_5}=T_2^{s_x} u^{-s_{\delta_2}}$,
and
\begin{align}
        \tilde{R_3} = &\hat{t}(T_3,g_2)^{s_x}\hat{t}(h, w)^{-s_{\alpha}-s_{\beta}}  \nonumber \\
                             &\hat{t}(h,g_2)^{-s_{\delta_1}-s_{\delta_2}}\left(\frac{\hat{t}(T_3, w)}{\hat{t}(g_1,g_2)}\right)^c. \nonumber
        \end{align}

The signature $\sigma$ is valid if $c=H(M, T_1, T_2, T_3,$ $\tilde{R_1},
\tilde{R_2}, \tilde{R_3}, \tilde{R_4}, \tilde{R_5})$.  Otherwise it is
invalid.

\item $\BOpen(M, \sigma, \gmsk, \gpk)$:

To open the signature, run $\BVerify(M, \sigma, \gpk)$.  If $\sigma$
is a valid signature on $M$, then the first part of the signer's secret
key can be retrieved: $A = \frac{T_3}{T_1^{\eta_1} T_2^{\eta_2}}$.
\end{itemize}

\subsection{An extension of the BBS Scheme}
\label{sub:BBSmod}

Suppose that every group member $b$ has some value in $\Z_p$ assigned
to it by the group manager.  This value changes with time, so that at
some time interval $\T_i$, this value is $r_{bi}$.  We want to modify the
$\BBS$ scheme in such a way that this value $r_{bi}$ is bound to the
group member's signature and is visible from it.  When $r_{bi}$
changes, the group member is able to obtain an update without a secure
channel.  The group public key $\gpk$ will also have to be modified
accordingly using some public information.  We will call this modified
scheme the $\BBS^*$ scheme, and it consists of the algorithms
$(\BKeyGen^*$, $\BJoin^*$, $\BUpdate^*$, $\BSign^*$, $\BVerify^*$,
$\BOpen^*)$.

\begin{itemize}
\item $\BKeyGen^*$:

In addition to the parameters generated in $\BKeyGen$, we have
the following public parameters:

  \begin{itemize}
  \item Time intervals $\T_0$, $\T_1$, $\T_2$, $\ldots$.
  \item For each time internal $\T_i$, we have a random base value
    $k_i \in \G_1 \setminus \{1_{G_1}\}$. A possible way to compute
    $k_i$ from $\T_i$ is using a public hash function, say $H'$, so
    that $k_i = H'(\T_i) \in G_1 \setminus \{1_{G_1}\}$.
  \item A set of values $\mathcal{R} = \{0, 1, 2, ..., m\} \subset
    \Z_p$, where $m < p$.  In each time interval $\T_i$ a group member
    $b$ has a specific value, denoted by $r_{bi} \in \mathcal{R}$
    assigned to it.
  \end{itemize}

 For each value of $r \in \mathcal{R}$, and each time interval $\T_i$
 we have a group public key denoted by $\gpk_{ir}$,
 \[ \gpk_{ir} = (\hat{g}_{1ir} = g_1 \cdot k_i^r, g_2,u,v,h,w). \]

 Hence we have $m+1$ group public keys $\gpk_{ir}$ in each time
 interval.  The secret key of the group manager is as before, $\gmsk =
 (\gamma, \eta_1, \eta_2)$.

\item $\BJoin^*(b, \gmsk)$:

This is the same as $\BJoin(b, \gmsk)$.
Each group member $b$ is given a secret key $\gsk_b=(A_b, x_b)$,
where $x_b \asn \Z_p^*$, and $A_b = g_1^{\frac{1}{\gamma+x_b}} \in \G_1$.

\item $\BUpdate^*(b,i, r_{bi}, \gsk_b, \gmsk)$:

At time interval $\T_i$, the group member $b$ which has value $r_{bi}$
may obtain an update of its secret signing key $\gsk_b=(A_b, x_b)$ as
follows.

The group manager computes $k_i = H'(\T_i)$, $R_{i} = k_i^{r_{bi}}$,
 $\rcert_{i} = R_i^{\frac{1}{\gamma + x_b}}$, and updates $A_b$ to $A_{bi}$ 
where $A_{bi} = A_b \cdot \rcert_{i}$.

The group member $b$ is given $\rcert_{i}$ publicly. When $b$ receives
$\rcert_{i}$ it first checks whether $\hat{t}(\rcert_i, wg_2^{x_b})
  =\hat{t}(R_i, g_2)$. If so, it then updates its secret signing key
$\gsk_b=(A_b, x_b)$ to $\gsk_{bi} = (A_{bi}, x_b)$; otherwise the
  received $\rcert_{i}$ is discarded (as it is corrupted or tampered
  with during the transmission).

\item $\BSign^*(M, i, r_{bi}, \gsk_{bi}, \gpk_{ir_{bi}})$:

To sign the message $M$ at time interval $\T_i$, a group member
$b$ with assigned value $r_{bi}$ performs
$\BSign(M, \gsk_{bi}, \gpk_{ir_{bi}})$.  The signature on
$M$ is $\sigma^{*} = (T_1,T_2,T_3, c, s_{\alpha}, s_{\beta},
s_{x}, s_{\delta_1}, s_{\delta_2}, i, r_{bi})$.

\item $\BVerify^*(M, \sigma^{*}, \gpk)$:

To verify the signature $\sigma^{*}$ on $M$, signed by a group member
with assigned value $r$ in the time interval $\T_i$,
i.e.\ $\sigma^{*}= (T_1,T_2,T_3, c, s_{\alpha}, s_{\beta}, s_{x},
  s_{\delta_1}, s_{\delta_2}, i, r)$, the verifier updates $\gpk$ to
$\gpk_{ir} = (\hat{g}_{1ir}, g_2,u,v,h,w)$ by computing
$\hat{g}_{1ir}= g_1 \cdot k_i^r$.  It then uses $\BVerify(M, \sigma,
\gpk_{ir})$ to verify if $\sigma$ is valid, where
  $\sigma=(T_1,T_2,T_3, c, s_{\alpha}, s_{\beta}, s_{x}, s_{\delta_1},
  s_{\delta_2})$.

\item $\BOpen^*(M, \sigma^{*}, \gmsk, \gpk)$:

To open the signature $\sigma$ on $M$, signed by a group member with
assigned value $r$ in the time interval $\T_i$, run $\BVerify^*(M,
\sigma^{*}, \gpk)$ first.  If the signature is valid then the first
part of the signer's secret key in time interval $\T_i$ can be
retrieved: $A_{bi} = \frac{T_3}{T_1^{\eta_1} T_2^{\eta_2}}$.

\end{itemize}

\subsection{Security of the $\BBS^*$ scheme}

We argue that the $\BBS^*$ scheme is both correct and secure. 

It is straightforward to verify that the $\BBS^*$ scheme is correct.
In fact, each instance of the $\BBS^{*}$ scheme is indeed
a $\BBS$ scheme. 

The modification of $\BBS$ to $\BBS^*$ consists of multiplying $g_1$
in the public key $\gpk$ with a public value $k_i^r$, sending
$\rcert_{i}$ publicly and using it to modify part of the user
$b$'s secret key $A_b$.  We argue that neither of these changes affect
the security of $\BBS$:

\begin{itemize}
\item {\bf Multiplying $g_1$ with a public value}: This does not
  affect the group manager's secret key and does not allow forgery of
  group members' secret keys.
\item {\bf Sending $\rcert_{i}$ publicly}: This does not reveal the
  secret values of $\gamma$, $A_{b}$ or $x_{b}$ if $\BBS$ is secure.
  If an adversary could obtain $\gamma$ or $x_{b}$ from $\rcert_i$
  then setting $R_i = g_1$, the adversary could also obtain $\gamma$
  or $x_{b}$ from $A_b$, thus allowing it to forge further group
  members' secret keys.
\end{itemize}

\section{Using $\BBS^*$ to enable a privacy-aware scheme}
\label{sec:multiple_description}

We now show how to deploy $\BBS^*$ to enable a privacy-aware
announcement scheme.
This scheme has a centralised architecture with two off-line central
authorities $\AS$, $\RS$, and \textit{vehicles} ($V$s) as end users.
The roles of these entities are as described in Section
\ref{sub:wivec_description}.  The management of the
reputation system is the same as the scheme of \cite{wivec}.

Let $\calR=\{0,1, \ldots, m\}$, $m < p$, represent the $m+1$
reputation levels.  At time interval $\T_i$, a vehicle $\Vb$
has a specific reputation level, denoted by $r_{bi}$.  The method
on how to establish such a level for a vehicle is the same as the
method used in the scheme of \cite{wivec}.  The group signature
scheme $\BBS^*$ allows the binding of the reputation level visibly
to a group signature.

Now we describe this new scheme in detail.  We will follow the same
presentation structure as used in Section \ref{sec:wivec}.

\subsection{Scheme Initialisation}
This is executed once only, when the announcement scheme is set up.
\begin{enumerate}
\item The $\AS$ regulates its clock, and deploys its remote wireless
  communication interfaces.

\item The $\RS$ creates a database, and installs $\Aggr$ and
  $\Detect$.

\item The $\AS$ installs $\mathsf{BBS^*}$, $\TimeDiscount$, and $\Psi$
  and divides the time into time intervals $(\T_{0}, \T_{1}, \T_{2},
  \cdots)$.

\item The $\AS$ executes $\BKeyGen^*$ to obtain $(\G_1, \G_2, \G_3,
  g_1, g_2, \psi, \hat{t}, H)$ and $\AS$'s public key is $\gpk$ and
  secret key is $\gmsk$.
\end{enumerate}

\subsection{Vehicle Registration}
This is executed when a new vehicle $\Vb$ requests to join the
announcement scheme.  It takes place in a secure environment: all
communication is confidential and authenticated.

\begin{enumerate}
\item The $\AS$ provides $V$ with $\BUpdate^*$, $\BSign^*$, $\BVerify^*$,
  and $\gpk$.

\item The $\AS$ and $\Vb$ executes $\BJoin^*(b, \gmsk)$, and $\Vb$
  recieves its group member secret key $\gsk_b = (A_b, x_b)$.

\item The $\AS$ requests the $\RS$ to create a record in its database
  for vehicle $\Vb$, indexed by $A_b$.
\end{enumerate}

\subsection{Reputation Retrieval}
When a vehicle $V_{b}$ drives into the proximity of a wireless
communication interface at time $\T_i$, it retrieves its reputation
information as follows:

\begin{enumerate}
\item $\Vb$ signs a reputation score request using $\gsk_{bi}$.  This
  authenticates $\Vb$ to $\AS$.  This signature is then opened using
  $\BOpen^*$ and $\AS$ is thus able to request the correct reputation
  score from $\RS$.

\item Upon retrieving $(r_{bi},\Vb, t_i)$, the reputation score of $\Vb$ at
  the current time $t_i$, from the $\RS$, the $\AS$ computes $\Vb$'s
  time discounted reputation scores $(r'_{i}, r'_{i+1}, \cdots,
  r'_{i+d})$ until $r'_{i+d+1} < \Psi$.

\item The $\AS$ then calculates $R_j = k_j^{r_{j}'}$ for public $k_j$
  and $\rcert_{j} = R_j^{\frac{1}{\gamma + x_b}}$ for $j =i, \ldots i+d$.

\item The $\AS$ sends $\rcert_{i}, \rcert_{i+1}, \ldots, \rcert_{i+d}$
  to $\Vb$ publicly and keeps a record of them.

\item $\Vb$ checks whether $\hat{t}(\rcert_j, wg_2^{x_b}) =
  \hat{t}(R_j, g_2)$ for $j =i, \ldots i+d$. If so then it updates its
  signing key $\gsk=(A_b, x_b)$ to $\gsk_{bj} = (A_{bj}, x_b)$ where
  $A_{bj} = A_b \cdot \rcert_{j}$, $j =i, \ldots i+d$.  In essence
  $\Vb$ and $\AS$ run $\BUpdate^*(b,j,r_{bj},\gsk_{bj},\gmsk)$ for $j
  =i, \ldots i+d$.

\end{enumerate}

\subsection{Message Broadcast}
A message $M$ is broadcast by $\Vb$ at time interval $\T_i$ as follows:

\begin{enumerate}
\item $\Vb$ retrieves the current time from its clock and identifies
  its corresponding time interval, say $\T_{i}$.

\item $V_{b}$ uses $\BSign^*(M, i, r_{bi}, \gsk_{bi},
  \gpk_{ir_{bi}})$ to generate a signature $\sigma^{*}$ on $M$, and forms
  a \textit{message tuple} $\msg = (M, \sigma^{*})$. $\Vb$ then
  broadcasts $\msg$ to its neighbouring vehicles.

\item Upon receiving $\msg= (M, \sigma^*)$, a receiving vehicle $\Vr$
  immediately identifies the current time interval $\T_{j}$ from its
  clock. $\Vr$ checks if $j = i$. If so then $\Vr$ uses $\BVerify^*(M,
  \sigma^{*}, \gpk)$ to verify $\sigma^{*}$. Upon successful
  verification, $\Vr$ can now decide whether to trust the announcement
  based on its own policy.  The message tuple $\msg$ is stored for
  future possible feedback reporting. If $j \not= i$ or the
  verification fails then $\Vr$ does not consider $V_{b}$ to be
  reputable, and thus discards $M$.
\end{enumerate}

\subsection{Feedback reporting}
When $\Vr$ has experience of the event described by message $M$, it is
able to judge the reliability of $M$. Then $\Vr$ can voluntarily
report feedback.

\begin{enumerate}
\item $\Vr$ assigns a feedback $f$ based on its experience about the
  reliability of $M$ and forms a feedback report $\fr=(f, \msg)$.
\item When $\Vr$ drives into the proximity of a wireless
  communication interface during time interval $\T_j$,
  $\Vr$ sends $\fr$ and $\BSign^*(\fr, j, r_{rj}, \gsk_{rj},
  \gpk_{jr_{rj}})$ to $\AS$.
\item The $\AS$ verifies $\Vr$'s signature.  If it is valid it
runs $\BOpen^*(\msg, \gmsk, \gpk)$ to obtain $A_{bi}$.  It then sends
the corresponding feedback $f$ to $\RS$.
\end{enumerate}

\subsection{Vehicle Revocation}
The $\AS$ revokes the identified malicious vehicle by no longer
providing them with new $\rcert_i$ in the future.  The revoked vehicle
will not be able to construct valid signatures without $\rcert_i$.

\subsection{Privacy and Robustness}
\label{sub:newpriv}

The privacy of this scheme, as in the scheme of \cite{wivec}, depends
on the security of $\MEA^+$ and $\TOA^+$.  If $\BBS^*$ is secure then
all data sent by a vehicle is protected with respect to anonymity and
unlinkability against all entites except for the $\AS$.

Observe that our privacy-aware scheme still features the same
robustness as the schemes of \cite{wivec,Li2012reputation} against
adversaries. An adversary is not able to impersonate an existing
vehicle or forge a legitimate broadcast message. This is because group
member signing keys are updated securely in $\BBS^*$ by legitimate
vehicles, and external adversaries are unable to obtain a valid group
member secret key. In addition, all approaches that can be used in
\cite{wivec,Li2012reputation} to prevent internal adversaries
conducting reputation manipulation can also be used in this new
scheme.

\subsection{A note on Computational and Communication Overheads}
\label{sub:overheads}

Group signatures are generally regarded as resource intensive and
time-consuming.  We briefly comment on the additional computational
and communication burden in using $\BBS^*$ for VANET announcements
compared to \cite{wivec}.

Firstly, there are VANET announcement schemes using the group signature scheme
$\BBS$  and they are shown to be feasible theoretically and by simulation,
for example, in \cite{GSIS,wivec}.
The new $\BBS^*$ scheme is based on $\BBS$, with 
a few more operations:
\begin{itemize}
\item $\BUpdate^*$ performs one check and one calculation.  
The check involves 2 pairings, 1 point multiplication and 1 
exponentiation.   The calculation requires 1 point multiplication.
\item $\BVerify*$ requires 1 additional point multiplication and 
1 exponentiation.
\end{itemize}
Altogether the $\BBS^*$ scheme requires 2 extra pairings, 3 extra
point multiplications and 2 extra exponentiation compared to
\cite{wivec}.  However, this is instead of having to establish a
secure and private channel for reputation retrieval, which requires
encryption as well as a digital signature.  For a vehicle to sign a
request and to verify a signature from the server will take 1 pairing,
2 point multiplications and 3 exponentiations (1 exponentiation for
signing, 2 exponentiation, 2 multiplication, 1 pairing for
verification) for the Boneh-Boyen scheme \cite{BonehBoyen}.  Hence the
computational overhead to being able to retrieve private values in
public is about 1 pairing and 1 point multiplication.

To be conservative, even for 128-bit security (most proposals are
using 80-bit security which is sufficient since most VANET
announcements are ephemeral) and using only 400 MHz processor
\cite{VSC}, 1 pairing will take 5 ms \cite{timing-pairings} and 1
multiplication will take 0.5 ms (using 200 000 cycle
per second for multiplication, which is also conservative according to
\cite{timing-mult}.)
Hence we add at most 5.5 ms.

As for signature length, we have two more elements, $i$ and $r_{bi}$.
We take 4 bytes for $i$ (a time-related parameter, 4 bytes are
sufficient for timestamps \cite{GSIS}) and 170 bits for $r_{bi}$
(which is an element of $\Z_p$, and $p$ is 170 bits for 80-bit
security).  This adds to the original $\BBS$ signature of length 1533
bits \cite{boneh2004short}, so we have a signature for $\BBS^*$ with
length 1735 bits (about 217 bytes), and this is under 250 bytes which
is the requirement for vehicular communications \cite{DSRC}.

The additional download for $\BUpdate^*$ is public and $\rcert_i$ is also
170 bits only, so this does  not present a barrier.

\section{Conclusion}
\label{sec:conclusion}

We have shown a reputation-based announcement scheme in VANETs which
supports flexible decision-making using explicit multiple reputation
levels - a vehicle may decide on its own policy whether to trust
announcements of different types depending on the announcing vehicle's
reputation score.  It also allows private reputation score retrieval
via a public channel, thus preserving user privacy across the wireless
interface.  This is enabled by our construction of a new primitive
based on a group signature scheme.  Two questions are of interest:

\begin{enumerate}
\item Can this privacy-aware reputation scheme be used for other types
  of network?  The robustness of this scheme against reputation
  manipulation depends on the relatively slow propagation of data.
  VANETs meet this requirement since data transmissions is largely
  achieved by short-range wireless medium.  How robustness can be
  acheived while guaranteeing privacy in a network with fast
  propagation, such as the internet, seems to be a hard problem.
\item Are there other applications for the primitive $\BBS^*$?  This
  offers a feature that allows a user to demonstrate some property
  within a group signature. In this particular application, the
  property is presented by two values, a time and a reputation
  score. In general, the property could be anything, such as a degree,
  a location or a position, and multiple properties can be bound
  together in one signature.  Similar ideas have been considered in
  other areas, such as anonymous credential and attribute-based
  signatures, and we believe $\BBS^*$ may turn out to be of
  independent interest.
\end{enumerate}

\end{document}